\documentclass[english,twocolumn,showpacs,superscriptaddress,prl]{revtex4}
\usepackage[T1]{fontenc}
\usepackage[latin9]{inputenc}
\usepackage{amsmath}
\usepackage{graphicx}
\usepackage{amssymb}
\usepackage{babel}

\begin{document}

\title{Spin field effect transistors with ultracold atoms}

\date{\today{}}

\author{J. Y. Vaishnav}

\affiliation{Joint Quantum Institute, National Institute of Standards and Technology,
Gaithersburg MD 20899 USA}

\author{Julius Ruseckas}

\affiliation{Institute of Theoretical Physics and Astronomy of Vilnius University,
A. Go\v{s}tauto 12, Vilnius 01108, Lithuania}

\author{Charles W. Clark }

\affiliation{Joint Quantum Institute, National Institute of Standards and Technology,
Gaithersburg MD 20899 USA}

\author{Gediminas Juzel\=unas}

\affiliation{Institute of Theoretical Physics and Astronomy of Vilnius University,
A. Go\v{s}tauto 12, Vilnius 01108, Lithuania}

\begin{abstract}
We propose a method of constructing cold atom analogs of the spintronic
device known as the Datta-Das transistor (DDT), which despite its
seminal conceptual role in spintronics, has never been successfully
realized with electrons. We propose two alternative schemes for an
atomic DDT, both of which are based on the experimental setup for
tripod stimulated Raman adiabatic passage. Both setups involve atomic
beams incident on a series of laser fields mimicking the relativistic
spin orbit coupling for electrons that is the operating mechanism
of the DDT.
\end{abstract}


\maketitle
\begin{figure}
\begin{centering}
\includegraphics[width=0.8\columnwidth]{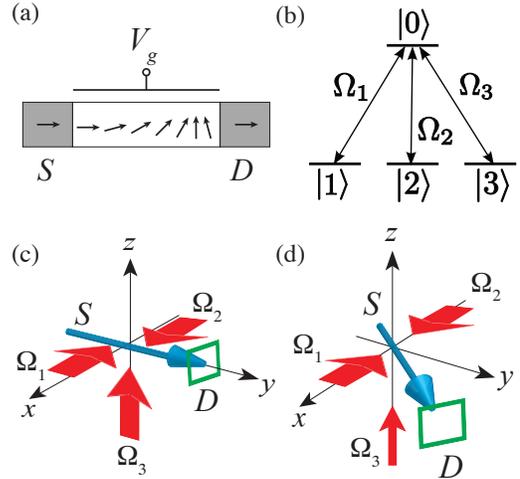}
\par\end{centering}

\caption{(a) Schematic of a DDT. {}``S'' and {}``D'' are ferromagnetic
source and drain electrodes. In between is a semiconducting gate region,
where the spin precesses by an amount which depends periodically on
the tunable gate voltage $V_{g}.$ This precession results in a controllable
current modulation at $D.$ (b) A tripod scheme of atomic energy levels,
coupled by laser fields with Rabi frequencies $\Omega_{i}$. (c,d)
Two alternative setups for an atomic version of the DDT. Here, the
source is a state-polarized atomic beam (blue), the gate is the intersection
region of a configuration of laser beams (red), and the drain is an
atomic state analyzer (green). }

\label{fig:ddtschematic}
\end{figure}

The emerging technology of semiconductor spintronics exploits the
electron's spin degree of freedom, as well as its charge state. The
first scheme for a semiconductor spintronic device was a spin field-effect
transistor known as the Datta-Das transistor (DDT) (Fig.~\ref{fig:ddtschematic}a)
\citep{dattadas}. The eighteen years since the theoretical proposal
have seen numerous experimental efforts to construct the DDT. Various
experimental obstacles, such as difficulties in spin injection, stray
electric fields and insufficient quality of spin-orbit coupling, have
prevented successful implementation of the DDT \citep{zutic2004sfa}.

Cold atom systems, in contrast with their electronic counterparts, are
highly controllable and tunable. This suggests the possibility of
designing precise atomic analogs of electronic systems which, due
either to fundamental physical limits or technological difficulties,
are experimentally inaccessible in their original manifestations.
The idea grows out of recent interest in {}``atomtronics,'' or building
cold atom analogs of ordinary electronic materials, devices and circuits
\citep{Atom-Diode,seaman:023615,experimentaldiode}. In particular,
an atom diode has been proposed \citep{Atom-Diode} and realized \citep{experimentaldiode}. 

In this Letter, we identify a method for constructing a cold atom
analog of a Datta-Das transistor. The setup is based on a four level
{}``tripod'' scheme of atom-light coupling \citep{unanyan1998rca,unanyan99pra,theuer1999nlc,stirap,ruseckas-2005-95,lietfiz-2007}
involving three atomic ground states and one excited state (see Fig.~\ref{fig:ddtschematic}b).
Such tripod schemes are an extension of the usual three-level $\Lambda$-type
setup for stimulated Raman adiabatic passage (STIRAP) \citep{Atom-Diode,spielman-2008},
and are experimentally accessible in metastable Ne, $^{87}\mbox{Rb}$
and a number of other gases \citep{theuer1999nlc,stirap}. The proposed
device provides a robust method for atomic state manipulation that
is immune to the inhomogeneities intrinsic to programmed Rabi pulses. 

The source terminal of an electronic DDT (Fig.~\ref{fig:ddtschematic}a)
is a ferromagnetic electrode that emits spin-polarized electrons.
The DDT drain terminal, a ferromagnetic analyzer, acts as a spin filter.
Between source and drain is a semiconducting gate region, in which
the gate-induced electric field produces a Rashba spin-orbit coupling
\citep{Rashba60} for electrons. While passing through the gate region,
the electron's spin precesses; the electron emerges at the drain having
undergone a spin rotation which is tunable via the gate voltage. Since
the drain passes only a certain spin direction, the drain current
is an oscillating function of the gate voltage.

Our atomic analog of the DDT (Figs.~\ref{fig:ddtschematic}c,d) uses
a beam of atoms in place of electrons. The two dark states in the
tripod setup play the role of the electron's spin states, and the
{}``source'' is a dilute atomic beam. The {}``gate'' region consists
of crossed laser beams engineered to mimic Rashba or Rashba-like spin
orbit couplings \citep{galitski,Jacob07,jayzb,prar08,juz08-neg-refl};
the analog of the gate voltage can be tuned by varying the relative
strengths of the lasers. The drain is a state-selective atomic filter,
such as a Stern-Gerlach device or radio-frequency or Raman outcoupler
\citep{Edwards}. While the goal of this paper is to explore the possibility
of constructing the atomic analog of spintronic devices, the two dark
states of the tripod atom can be considered qubit states \citep{qi2,qi0,unanyan04PRA,qi3};
in this context the atomic DDT represents a single-qubit phase gate
for a dilute atomic beam. In contrast to typical single qubit gates,
this setup does not involve time-dependent pulses, and the amount
of the qubit rotation within the gate region is independent of the
atom's velocity, due to the geometric nature of the process.

\paragraph{Tripod scheme}

The proposed DDT implementations exploit the tripod scheme (Fig.~\ref{fig:ddtschematic}b,c)
\citep{unanyan1998rca,unanyan99pra,theuer1999nlc,stirap,ruseckas-2005-95,lietfiz-2007},
in which a four level atom feels two counterpropagating stationary
laser beams and a third orthogonal beam \citep{Jacob07,prar08,juz08-neg-refl}.
The lasers induce transitions between the ground states $|j\rangle$
($j=1,2,3$) and an excited state $|0\rangle$ with spatially dependent
Rabi frequencies $\Omega_{1}=|\Omega_{1}|e^{-i\kappa_{0}x}$ , $\Omega_{2}=|\Omega_{2}|e^{i\kappa_{0}x}$
and $\Omega_{3}=|\Omega_{3}|e^{i\kappa_{0}z}$ , $\kappa_{0}$ being
a wave-number.

The electronic Hamiltonian of a tripod atom is, in the interaction
representation and rotating wave approximation, $\hat{H}_{e}=-\hbar\Omega|B\rangle\langle0|+\mathrm{H.c.}$,
where $|B\rangle=\left(|1\rangle\Omega_{1}^{*}+|2\rangle\Omega_{2}^{*}+|3\rangle\Omega_{3}^{*}\right)/\Omega$
and $\Omega^{2}=|\Omega_{1}|^{2}+|\Omega_{2}|^{2}+|\Omega_{3}|^{2}$.
$\hat{H}_{e}$ has two degenerate dark states $|D_{j}\rangle$ containing
no excited state contribution: $\hat{H}_{e}|D_{j}\rangle=0$, $j=1,2.$
An additional pair of bright eigenstates $|\pm\rangle=\left(|B\rangle\pm|0\rangle\right)/\sqrt{2}$
is separated from the dark states by $\pm\hbar\Omega$. For the light
fields of interest, the dark states can be chosen as:

\begin{eqnarray}
|D_{1}\rangle & = & \left(\sin\varphi|1\rangle^{\prime}-\cos\varphi|2\rangle^{\prime}\right)\,,\label{eq:D1}\\
|D_{2}\rangle & = & \varepsilon\left(\cos\varphi|1\rangle^{\prime}+\sin\varphi|2\rangle^{\prime}\right)-\sqrt{1-\varepsilon^{2}}|3\rangle,\label{eq:D2}\end{eqnarray}
with $|1\rangle^{\prime}=|1\rangle e^{i\kappa_{0}(z+x)}$ and $|2\rangle^{\prime}=|2\rangle e^{i\kappa_{0}(z-x)}$,
where\begin{equation}
\varepsilon=|\Omega_{3}|/\Omega\,,\quad\varphi=\arctan(|\Omega_{1}|/|\Omega_{2}|)\label{eq:epsilon-phi}\end{equation}
characterize the relative intensities of the laser beams. The dark
states $|D_{j}\rangle\equiv|D_{j}(\mathbf{r})\rangle$ are position-dependent
due to the spatial variation of the Rabi frequencies $\Omega_{j}(\mathbf{r})$.

Let us adiabatically eliminate the bright states, so that the atom
evolves within the dark-state manifold.  The full atomic state vector
can then be expanded as $|\Psi(\mathbf{r},t)\rangle=\sum_{n=1}^{2}\chi_{n}(\mathbf{r},t)|D_{n}(\mathbf{r})\rangle$,
where $\chi_{n}(\mathbf{r},t)$ describes the motion of an atom in
the dark state $|D_{n}(\mathbf{r})\rangle$. The atomic center of
mass motion is thus represented by a two-component wavefunction $\chi=(\chi_{1},\chi_{2})^{T}$
obeying \citep{ruseckas-2005-95}\textbf{\begin{equation}
i\hbar\frac{\partial}{\partial t}\chi=\left[\frac{1}{2M}(-i\hbar\nabla-\mathbf{A})^{2}+U\right]\chi\,,\label{eq:general-eq}\end{equation}
}where $\mathbf{A}$ is the effective vector potential \citep{ruseckas-2005-95,berry84,wilczek84,mead91}
representing a $2\times2$ matrix whose elements are vectors, $\mathbf{A}_{n,m}=i\hbar\langle D_{n}(\mathbf{r})|\nabla D_{m}(\mathbf{r})\rangle$.
The particular light field configuration we have chosen yields\begin{eqnarray}
\mathbf{A}_{11} & = & -\hbar\kappa_{0}(\mathbf{e}_{z}-\cos(2\varphi)\mathbf{e}_{x}),\label{eq:A-11}\\
\mathbf{A}_{12} & = & -\hbar\varepsilon(\kappa_{0}\sin(2\varphi)\mathbf{e}_{x}+i\nabla\varphi),\label{eq:A-12}\\
\mathbf{A}_{22} & = & -\hbar\kappa_{0}\varepsilon^{2}(\mathbf{e}_{z}+\cos(2\varphi)\mathbf{e}_{x}),\label{eq:A-22}\end{eqnarray}
with $\mathbf{e}_{x}$ and $\mathbf{e}_{z}$ the unit Cartesian vectors.
The $2\times2$ matrix $U$ with elements $U_{nm}=(\hbar^{2}/2M)\langle D_{n}(\mathbf{r})|\nabla B(\mathbf{r})\rangle\langle B(\mathbf{r})|\nabla D_{m}(\mathbf{r})\rangle$
is an effective scalar potential; both $\mathbf{A}$ and $U$ arise
due to the spatial dependence of the atomic dark states. 

Suppose the incident atom has a velocity $\mathbf{v}$ much greater
than the recoil velocity $v_{\mathrm{rec}}=\hbar\kappa_{0}/M\approx0.5\mbox{cm/s}$
for $^{87}$Rb. In this limit, the laser beams do not significantly
change the atom's velocity, permitting a simplified semiclassical
approach with no reflected waves. We apply a gauge transformation
$\chi(\mathbf{r},t)=e^{iM\mathbf{v}\cdot\mathbf{r}/\hbar-iM\mathbf{v}^{2}t/2\hbar}\tilde{\chi}(\mathbf{r},t)$,
implying transition to a reference frame moving with velocity $\mathbf{v}$,
where the two-component envelope function $\tilde{\chi}$ varies slowly
with $\mathbf{r}$ over the atom's wavelength $\lambda=h/(Mv)$. Keeping
only terms containing $\mathbf{v}$ (or its time derivatives), we
arrive at the following approximate equation for $\tilde{\chi}$:

\begin{equation}
i\hbar\left(\partial/\partial t+\mathbf{v}\cdot\nabla\right)\tilde{\chi}(\mathbf{r},t)=-\mathbf{v}\cdot\mathbf{A}(\mathbf{r})\tilde{\chi}(\mathbf{r},t).\label{eq:specific-eq-envelope}\end{equation}
As the omitted scalar potential $U$ and the $A^{2}$ term are of
the order of the recoil energy $\hbar\omega_{\mathrm{rec}}=\hbar^{2}\kappa_{0}^{2}/2M\ll Mv^{2}/2$,
the fast moving atoms will not feel these potentials. For incident
velocities $v$ of the order of $v_{\mathrm{rec}}$ or smaller, the
atomic motion will undergo a \emph{Zitterbewegung} \citep{jayzb,patrikzb}
which is beyond the scope of the present study. While the atoms must
move much faster than the recoil velocity, they should also be slow
enough to avoid coupling to the bright states. We provide a quantitative
analysis of these limitations near the end of the Letter.

In both of the DDT schemes to be presented, the operator \textbf{$\mathbf{v}\cdot\mathbf{A}$}
commutes with itself at different times. Going to a moving frame of
reference $\mathbf{r}^{\prime}=\mathbf{r}-\mathbf{v}t$, we can thus
relate the wavefunction $\tilde{\chi}$ at time $t=t_{f}$ to the
wavefunction at a previous time $t=t_{i}$ through\begin{equation}
\tilde{\chi}(\mathbf{r}^{\prime},t_{f})=\exp(i\Theta)\tilde{\chi}(\mathbf{r}^{\prime},t_{i})\,.\label{eq:envelope-solution}\end{equation}
The $2\times2$ Hermitian matrix $\Theta=-\hbar^{-1}\int_{t_{i}}^{t_{f}}\mathbf{A}(\mathbf{r}^{\prime}+\mathbf{v}t)\cdot\mathbf{v}dt$
describes the evolution of the internal state of the atom as it traverses
the path from $\mathbf{r}_{i}=\mathbf{r}^{\prime}+\mathbf{v}t_{i}$
to $\mathbf{r}_{f}=\mathbf{r}^{\prime}+\mathbf{v}t_{f}$, \begin{equation}
\Theta=-\frac{1}{\hbar}\int_{\mathbf{r}_{i}}^{\mathbf{r}_{f}}\mathbf{A}(\mathbf{r})\cdot d\mathbf{r}\,.\label{eq:theta}\end{equation}
Our subsequent analysis of the atomic dynamics will center on Eqs.
(\ref{eq:envelope-solution})-(\ref{eq:theta}) and (\ref{eq:A-11})-(\ref{eq:A-22}).

\paragraph{Atomic analogs of the DDT}

\begin{figure}
\begin{centering}
\includegraphics[width=0.75\columnwidth]{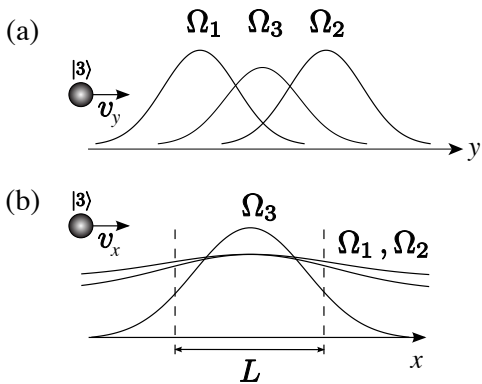}
\par\end{centering}

\caption{Schematics of the first (a) and second (b) setups for an atomic transistor:
The atom, along its trajectories (shown in Figs.~\ref{fig:ddtschematic}c,d)
sees the above profile of laser fields. }

\label{fig:razmik}
\end{figure}

We first consider the setup depicted in Figs.~\ref{fig:ddtschematic}c
and \ref{fig:razmik}a. The atoms are incident along the $y$ axis,
along which laser beams $1$ and $2$ are relatively shifted \citep{unanyan1998rca,theuer1999nlc,unanyan99pra,lietfiz-2007},
so that \begin{equation}
A_{y}=\hbar\sigma_{y}\varepsilon(y)\partial\varphi(y)/\partial y\,.\label{eq:A-y}\end{equation}
 Equations (\ref{eq:A-y}) and (\ref{eq:theta}) yield \begin{equation}
\Theta=\alpha\sigma_{y}\,,\qquad\alpha=-\int_{y_{i}}^{y_{f}}\varepsilon(y)\frac{\partial}{\partial y}\varphi(y)dy\,,\label{eq:alpha-Razmik}\end{equation}
where $\alpha$ is the mixing angle, $\sigma_{y}$ (or $\sigma_{x}$)
being the usual Pauli matrix. By taking the initial and final times
sufficiently large, we have $y_{i}\rightarrow-\infty$ and $y_{f}\rightarrow+\infty$. 

\begin{figure}
\begin{centering}
\includegraphics[width=0.55\columnwidth]{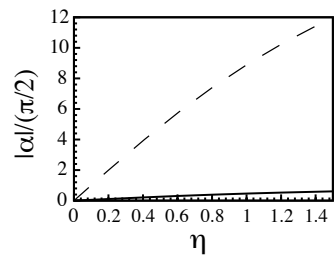}
\par\end{centering}

\caption{The mixing angle $\alpha$ vs. the relative amplitude of the third
field $\eta$ for the first (solid line) and the second (dashed line)
setups. The amplitudes of the beams are Gaussian: \textbf{$|\Omega_{1}|=a\exp(-(u+\delta)^{2}/w_{1}^{2})$},
\textbf{$|\Omega_{2}|=a\exp(-(u-\delta)^{2}/w_{2}^{2})$}, and \textbf{$|\Omega_{3}|=a\eta\exp(-u^{2}/w_{3}^{2}-\delta^{2}/w_{1}^{2})$},
with $u=y$ (first setup) or $u=x$ (second setup). In the first setup,
$w_{1}=w_{2}=w_{3}=\delta=2\lambda$, with $\lambda=600\,\mathrm{nm}$
being the laser wave length. For the second setup, all the beams are
centered at the same point ($\delta=0$) and have the widths $w_{1}=w_{2}=10w_{3}=20\lambda$.
\textbf{\label{fig:fig3}}}

\end{figure}

As Figs.~\ref{fig:ddtschematic}c and \ref{fig:razmik}a show, the
first laser beam dominates as the atom enters the gate region, while
the second dominates as it exits the region. In between, the atom
also feels the third beam. This configuration results in a gate-induced
rotation of the atom's internal state by a mixing angle \textbf{$\alpha$}.
Specifically, suppose the atom enters the gate region in the internal
state $|3\rangle=-|D_{2}(\mathbf{r}^{\prime},t_{i})\rangle$, with
center of mass wave-function \textbf{$\Phi(\mathbf{r}^{\prime})$}.
The atom then exits the gate region in the rotated state \begin{equation}
\tilde{\chi}(\mathbf{r}^{\prime},t_{f})=-\Phi(\mathbf{r}^{\prime})\left(\begin{array}{c}
\sin\alpha\\
\cos\alpha\end{array}\right).\label{eq:Psi-fin}\end{equation}
Thus, the probability for the atom to emerge in the second dark state
is \textbf{$\cos^{2}\alpha$}. Note that the second dark state coincides
with the third internal ground state upon exit:  \textbf{$|D_{2}(\mathbf{r}^{\prime},t_{f})\rangle=-|3\rangle$}.  This gate-controlled state rotation is an atomic analog of the action
of the DDT. Define $\eta=|\Omega_{3}|/|\Omega_{1}|$ as the relative amplitude of the third laser
at the central point. The specific relation between
$\alpha$ and $\eta$ depends on the particular choice of light field
configuration and is readily derived from Eqs. (\ref{eq:epsilon-phi})
and (\ref{eq:alpha-Razmik}).  For arbitrary light field configurations, $\alpha$ is a
complicated space-dependent function.  However for the particular
laser configuration we examine here, $\alpha$ simplifies to a function solely depending on $\eta$, and  $\eta$ controls $\alpha$.  Fig. \ref{fig:fig3}\textbf{ }shows the dependence of $\alpha$ on $\eta$ for Gaussian laser beams. As in the electronic DDT, the transmission coefficient $\cos\alpha$
is independent of the velocity of the incident atoms, so that the
transistor properties are robust to a spread in atomic velocities.
We estimate the regime of validity of this independence near the end
of the Letter.

Since $\varepsilon(y)\leq1$, the mixing angle given by Eq.~(\ref{eq:alpha-Razmik})
ranges from $0$ to $\pi/2$, and the sensitivity $|\Delta\alpha|/|\Delta\eta|$
of the DDT is on the order of unity. Small changes in the relative
Rabi frequency $\eta$ will thus lead to small changes in the mixing
angle: $|\Delta\alpha|\sim|\Delta\eta|$. We next analyze an alternative
setup which enables us to create a more sensitive DDT.

Now suppose that the first two light beams counterpropagate along
the $x$ axis with equal intensities (Fig.~\ref{fig:ddtschematic}d),
i.e., $\varphi=\pi/4$ in Eqs.~(\ref{eq:A-11})-(\ref{eq:A-22})
for $\mathbf{A}$. After the trivial gauge transformation $\exp[i\hbar\kappa_{0}(1+\varepsilon^{2})z\mathbf{I}]$,
the light-induced vector potential resembles the Rashba spin orbit
coupling which is the spin rotation mechanism of the electronic DDT: 

\begin{eqnarray}
A_{z} & = & -\frac{\hbar\kappa_{0}}{2}(1-\varepsilon^{2})\sigma_{z}\label{eq:A-2-setup}\\
A_{x} & = & -\hbar\kappa_{0}\varepsilon\sigma_{x},\qquad A_{y}=0.\end{eqnarray}

The atomic beam crosses the lasers at an angle in the $x-y$ plane,
with initial velocity components $v_{x}\neq0$ and $v_{y}$. Although
the atomic motion in the $y$ direction does not affect the internal
state rotation \textbf{$(A_{y}=0)$} , sending the beam in at an angle
removes the experimental difficulty of having the atoms incident from
the same direction as the laser beams. Along its trajectory, the atom
feels the laser beam profile illustrated in Fig.~\ref{fig:razmik}b.
The evolution matrix of Eq.~(\ref{eq:theta}) is then\begin{equation}
\Theta=\alpha\sigma_{x}\,,\qquad\alpha=\kappa_{0}\int_{x_{i}}^{x_{f}}\varepsilon(x)dx\,.\label{eq:theta-our}\end{equation}
Initial and final times are taken sufficiently large that the spatial
integration runs from $x_{i}=-\infty$ to $x_{f}=+\infty$.

As in the previous scheme, the intensity of the third laser vanishes
($\varepsilon\rightarrow+0$) outside the gate region (see Fig.~\ref{fig:razmik}b).
Only the third laser's intensity has significant spatial dependence
inside the gate region, the intensities of the first two lasers being
nearly constant there. In both setups, the controlled state rotation
arises from the spatial dependence of the beams in the gate region.
In the first setup the variation is in the lasers' relative intensities.
Contrastingly, in the second setup, the intensities of the first two
lasers are constant in the gate region, so the controlled state rotation
is driven by only the relative \emph{phases} of the counterpropagating
laser beams. 

As in the previous setup, the atom enters the gate region in the internal
state \textbf{$|3\rangle=-|D_{2}(\mathbf{r}^{\prime},t_{i})\rangle$}
and with center of mass wave-function $\Phi(\mathbf{r}^{\prime})$.
The atom exits in the rotated state\begin{equation}
\tilde{\chi}(\mathbf{r}^{\prime},t_{f})=-\Phi(\mathbf{r}^{\prime})\left(\begin{array}{c}
i\sin\alpha\\
\cos\alpha\end{array}\right),\label{eq:Phi-fin-altern}\end{equation}
where the mixing angle $\alpha$ is controlled by the variation of
the relative intensity of the third laser beam.

To estimate the mixing angle, suppose that $\Omega_{3}$, and hence
$\varepsilon$, do not change significantly in the gate region. Equation
(\ref{eq:theta-our}) then gives $\alpha=\kappa_{0}\bar{\varepsilon}L$,
where $L$ (see Fig.~\ref{fig:razmik}b) is the length of the area
in which the third laser has the strongest intensity. Note that the
mixing angle is now proportional to $L$, as well as to the average
strength $\kappa_{0}\bar{\varepsilon}$ of the spin-orbit coupling.
This behavior is in direct analogy to the electronic DDT \citep{dattadas}.
As in Eq.~(2) of \citep{dattadas}, the output power of the atoms
in the internal state $|3\rangle$ is $P=\cos^{2}\alpha=\cos^{2}(\kappa_{0}\bar{\varepsilon}L)$.
Using this atomic setup, $\alpha=\kappa_{0}\bar{\varepsilon}L$ can
be much larger than $\pi/2$, provided $L\gg(\kappa_{0}\bar{\varepsilon})^{-1}$,
as shown in Fig.~\ref{fig:fig3}. Small changes in the relative amplitude
of the third laser $\eta=|\Omega_{3}|/|\Omega_{1}|$ can therefore
yield substantial changes in the mixing angle: $|\Delta\alpha|\sim|\Delta\eta|\kappa_{0}L$.
The sensitivity of such a DDT, $|\Delta\alpha|/|\Delta\eta|\sim\kappa_{0}L$,
can far exceed unity if $L$ is much greater than the optical wave-length
$\lambda=2\pi/\kappa_{0}$. 

Let us estimate the range of atomic beam velocities for which our
approximations are valid. The atom crosses the gate region in a time
$\tau=L/v$. Due to nonadiabatic coupling to the bright states, the
dark state atoms have the finite lifetime $\tau_{D}=\Omega^{2}/\gamma\Delta\omega^{2}$
\citep{Juz06PRA}, where $\gamma$ is the excited state decay rate
and $\Delta\omega=v\partial\varphi/\partial y\sim v\pi/L$ (first
setup), or $\Delta\omega=v\kappa_{0}$ (second setup). The frequency
shift $\Delta\omega$ represents the two-photon detuning due to the
finite time of the atom-light interaction (first setup) or the two-photon
Doppler shift (second setup). To avoid decay, we require the beam
to be in the adiabatic limit, i.e. $\tau/\tau_{D}\ll1$. Taking $\Omega=2\pi\times10^{7}\,\mathrm{Hz}$
\citep{Hau99}, $\gamma=10^{7}\,\mathrm{s}^{-1}$, $\kappa_{0}=2\pi/\lambda$,
$\lambda=600\,\mathrm{nm}$ and $L=4\lambda$, we require atomic velocities
$v\ll100\,\mathrm{m/s}$ for the first setup and $v\ll1\,\mathrm{m/s}$
for the second setup. The increased sensitivity in the second scheme
thus comes at the expense of increased non-adiabatic losses.

Ultracold atoms are highly tunable and controllable, and can thus
serve as quantum simulators for a variety of other systems, including
systems which have yet to be experimentally accessed in their original
manifestations. In this Letter, we have identified an atomic analog
of one such system, the spin field-effect transistor. Our atomic transistors,
like their electronic counterpart, provide controllable state manipulation
that is relatively insensitive to the thermal spread of beam velocities.
The devices we have proposed are based on the familiar tripod STIRAP
configuration, and appear to be feasible within current experimental
procedures.

\bibliographystyle{prsty}

\begin{thebibliography}{10}

\bibitem{dattadas}
S. Datta and B. Das, Appl. Phys. Lett. {\bf 56},  665  (1990).

\bibitem{zutic2004sfa}
I. Zutic, J. Fabian, and S. Das~Sarma, Rev. Mod. Phys. {\bf 76},  323  (2004).

\bibitem{Atom-Diode}
A. Ruschhaupt, J.~G. Muga, and M.~G. Raizen, J. Phys. B: At. Mol. Opt. Phys.
  {\bf 39},  L133  (2006).

\bibitem{seaman:023615}
B.~T. Seaman, M. Kr\"{a}mer, D.~Z. Anderson, and M.~J. Holland, Phys. Rev. A
  {\bf 75},  023615  (2007).

\bibitem{experimentaldiode}
J.~J. Thorn, E.~A. Schoene, T. Li, and D.~A. Steck, Phys. Rev. Lett. {\bf 100},
   240407  (2008).

\bibitem{unanyan1998rca}
R.~G. Unanyan, M. Fleischhauer, B.~W. Shore, and K. Bergmann, Opt. Commun. {\bf
  155},  144  (1998).

\bibitem{unanyan99pra}
R.~G. Unanyan, B.~W. Shore, and K. Bergmann, Phys. Rev. A {\bf 59},  2910
  (1999).

\bibitem{theuer1999nlc}
H. Theuer {\it et~al.}, Opt. Express {\bf 4},  77  (1999).

\bibitem{stirap}
F. Vewinger {\it et~al.}, Phys. Rev. Lett. {\bf 91},  213001  (2003).

\bibitem{ruseckas-2005-95}
J. Ruseckas, G. Juzeli{\=u}nas, P. {\"O}hberg, and M. Fleischhauer, Phys. Rev.
  Lett. {\bf 95},  010404  (2005).

\bibitem{lietfiz-2007}
G. Juzeli{\=u}nas, J. Ruseckas, P. {\"O}hberg, and M. Fleischhauer, Lithuanian
  J. Phys {\bf 47},  351  (2007).

\bibitem{spielman-2008}
Y.-J. Lin {\it et~al.}, arXiv:0809.2976  (2008).

\bibitem{Rashba60}
E.~I. Rashba, Sov. Phys. Sol. St. {\bf 2},  1224  (1960).

\bibitem{galitski}
T.~D. Stanescu, C. Zhang, and V. Galitski, Phys. Rev. Lett. {\bf 99},  110403
  (2007).

\bibitem{Jacob07}
A. Jacob, P. {\"O}hberg, G. Juzeli{\=u}nas, and L. Santos, Appl. Phys. B {\bf
  89},  439  (2007).

\bibitem{jayzb}
J.~Y. Vaishnav and C.~W. Clark, Phys. Rev. Lett. {\bf 100},  153002  (2008).

\bibitem{prar08}
G. Juzeli{\=u}nas {\it et~al.}, Phys. Rev. A {\bf 77},  011802(R)  (2008).

\bibitem{juz08-neg-refl}
G. Juzeli{\=u}nas {\it et~al.}, Phys. Rev. Lett. {\bf 100},  200405  (2008).

\bibitem{Edwards}
M. Edwards {\it et~al.}, J. Phys. B: At. Mol. Opt. Phys. {\bf 32},  2935
  (1999).

\bibitem{qi2}
L.~M. Duan, J.~I. Cirac, and P. Zoller, Science {\bf 292},  1695  (2001).

\bibitem{qi0}
Z. Kis and F. Renzoni, Phys. Rev. A {\bf 65},  032318  (2002).

\bibitem{unanyan04PRA}
R.~G. Unanyan and M. Fleischhauer, Phys. Rev. A {\bf 69},  050302(R)  (2004).

\bibitem{qi3}
S. Rebi{\'c} {\it et~al.}, Phys. Rev. A {\bf 70},  032317  (2004).

\bibitem{berry84}
M.~V. Berry, Proc. R. Soc. A {\bf 392},  45  (1984).

\bibitem{wilczek84}
F. Wilczek and A. Zee, Phys. Rev. Lett. {\bf 52},  2111  (1984).

\bibitem{mead91}
C.~A. Mead, Rev. Mod. Phys. {\bf 64},  51  (1992).

\bibitem{patrikzb}
M. Merkl, F.~E. Zimmer, G. Juzeli{\=u}nas, and P. {\"O}hberg, Europhys. Lett.
  {\bf 83},  54002  (2008).

\bibitem{Juz06PRA}
G. Juzeli{\=u}nas, J. Ruseckas, P. \"Ohberg, and M. Fleischhauer, Phys. Rev. A {\bf 73},
  025602  (2006).

\bibitem{Hau99}
L. Hau, S.~E. Harris, Z. Dutton, and C. Behrooz, Nature {\bf 397},  594
  (1999).

\end{thebibliography}

\end{document}